\documentclass[11pt]{revtex4}
\usepackage{amssymb,epsf}
\usepackage{latexsym}
\usepackage{amsmath}
\begin{document}

\title{A new look at dark energy}
\author{N. Riazi$^{1}$\footnote{email: n\_riazi@sbu.ac.ir}
and  Sh. Assyyaee$^{2}$\footnote{email: s\_assyyaee@sbu.ac.ir}}
\affiliation{$1,2$. Physics Department, Shahid Beheshti
University, Tehran 19839, Iran.}
\begin{abstract}
The origin of the dark energy which is assumed to be responsible
for the observed accelerated expansion of the universe still
remains a scientific dilemma. Here we propose a tentative origin for this energy, if it is coming from a distribution of specific quantum particles.\\
\end{abstract}
\maketitle

Observational evidence for the accelerated expansion of the
universe \cite{0},\cite{1} demands the existence of a cosmic fluid
--called dark energy-- with unusual equation of state $p=w\rho$,
$w\sim -1$, if the general theory of relativity is valid on
cosmological scales\cite{wein}. Ordinary relativistic free
particles are known to yield a positive value for $w$. Although
the zero point energy of quantum fields nicely yield the required
value for $w$ (i.e. $w=-1$), the estimated value of the energy
density is vastly greater than the observed value\cite{wein2}. In
order to explain the value of dark energy density one is forced to
go beyond conventional explanations. In this letter, we look for
an explanation for dark energy in terms of quantum particles which
might exist in the Dirac energy gap. P.A.M. Dirac proposed in 1930
that there are infinitely many occupied states of any fermion
field with negative energies with $E\le -m$. Dirac postulated this
to explain the anomalous negative-energy quantum states predicted
by the Dirac equation for relativistic electrons. Although Dirac's
original proposal is not popular in modern quantum field theory, a
new interpretation has appeared in the theory of causal fermion
systems\cite{caus}. Since real particles have energies $E\ge m$
and cannot lead to negative pressure, in this letter we employ the
idea of particles which have energies within the forbidden Dirac
gap $-m< E < m$. Such particles should obviously have imaginary
momentum. Although unacceptable at the classical level, imaginary
momentum appear here and there in the quantum world in phenomena
like quantum tunneling\cite{raz}. Though to be more careful one
needs to distinguish the delicate difference between wave group
velocity and particle velocity to avoid the misinterpretation of
superluminal behavior of photons crossing the forbidden band
\cite{9,10}, still the tunnelling works well for our case, and
simultaneously it provides us a very nit explanation even for some
instrumental devises \cite{11,12}. The connection between
tunnelling and cosmological inflation previously discussed in
Wheeler-DeWitt equation footing \cite{13} or in a more modern
interpretation of quantum cosmology \cite{14}.

In order to make a link between quantum particles and the equation of state of the cosmic fluid, we start with the common relativistic expression  for the
energy-momentum tensor in terms of distribution function of particles\cite{6};

\begin{equation}\label{EMT}
T^{\mu\nu} = \int {{p^\mu p^\nu \over p^0} f(p)\mathrm{d^3}p},
\end{equation}
where $f(p)$ is the distribution function of particles with
4-momentum $p^\mu$ and $p_\mu p^\mu=m^2=E^2-p^2$ and in all
subsequent materials we respect the $(+,-,-,-)$ signature. The
energy momentum of a statistically homogenous distribution of
particles is given by \cite{6}:
\begin{equation}\label{EMTH}
T^{\mu\nu}=\sum_{i=1} ^N \int
{\delta^{(4)}[x-x_i(s)]{p_i}^\mu(s){d\over ds}{x_i}^\nu(s)}
\end{equation}
in which $x_i(s),i=1,...,N$ is the coordinate of the $i^{th}$
particle, ${p_i}^\mu(s)$ is the corresponding 4-momentum and $s$
is an invariant coordinate  along the particle geodesic. Equation
(\ref{EMTH}) can be put in the following form:
\begin{equation}\label{EMTR}
T^{\mu\nu}=\int{{{p^\mu p^\nu} \over m}R(p)\mathrm{d^4}p} ,
\end{equation}
where
\begin{equation}
R(p)=\int{{\sum_ {i=1}}^N
\delta^{(4)}[p-{p_i}(s)]\delta^{(4)}[x-x_i(s)]\mathrm{d}s} .
\end{equation}
The statistical distribution function is defined as
\begin{equation}
f(p)\equiv <R(p)>
\end{equation}
in which the average is taken over the initial data \cite{6}. One
can replace the integration element $d^4p$ with the three
dimensional element ${m \over p^0}d^3p$ to obtain:
\begin{equation}\label{EMT}
T^{\mu\nu} = \int {{p^\mu p^\nu \over p^0} f(p)\mathrm{d^3}p},
\end{equation}
 Since we are using a cosmological application of
(\ref{EMT}), it is a valid assumption to match (\ref{EMT}) with a
homogeneous and isotropic perfect fluid in complete accordance
with the cosmological principle and write;
\begin{equation}\label{dens}
T^0_0=\int {p^0 f(p)\mathrm{d^3}p}=\rho
\end{equation}
and
\begin{equation}\label{T11}
-P=T^1_1=\int {{p^1 p_1 \over E}
f(p)\mathrm{d^3}p}=\frac{1}{3}\int {{-p_x ^2 -p_y ^2 -p_z ^2 \over
E}f(p)\mathrm{d^3}p}=\int{{{m^2-E^2}\over E}f(p)\mathrm{d^3}p}
\end{equation}
Then in terms of perfect fluid density and pressure, this can be
written as;
\begin{equation}\label{Pr}
-P={g\over 3}\int {{f(p)\over E}\mathrm{d^3}p}-{1\over 3}\int
{Ef(p)\mathrm{d^3}p}
\end{equation}
which by (\ref{dens}) can be summarized as:
\begin{equation}
P=\frac{1}{3}\rho-{m^2\over 3}\int {{f(p)\over E}\mathrm{d^3}p}
\end{equation}
One can check that the above relation gives the correct answer for
radiation i.e. $(m=0)$ $P_{rad}=1/3 \rho_{rad}$ \cite{wein}. Now let
us determine the desired condition to have dark energy. It is well
known in cosmology that the observed acceleration of the universe
might come from a dark energy component with the equation of state
$P=w\rho$, $w<-1/3$. From (\ref{Pr}) we have:
\begin{equation}
P={m^2\over 3}\int {(E-{m^2\over E})f(p)\mathrm{d^3}p}
\end{equation}
To retrieve the negative pressure as a loose condition of dark
energy one requires the integrand to be negative:
\begin{equation}
E-{m^2\over E}<0
\end{equation}
which requires either
\begin{equation}\label{inequalities}
p^2<0 \quad or \quad E<0.
\end{equation}
In other words one must have either imaginary momentum or negative
energy.
So far we have not discussed the role of the distribution function
$f(p)$. The situation for relativistic particle in equilibrium is
well known. From statistical mechanics we have

\begin{equation}
f(p)={g \over h^3}{1\over e^{E/kT}\pm 1}
\end{equation}
In which g is the number of spin states and the $+(-)$ signs
correspond to fermions (bosons). However, for free, real particle in
equilibrium neither of the inequalities (\ref{inequalities}) are
satisfied. Therefore we have to go beyond the conventional
equilibrium distribution function, if we are to obtain something
like dark energy.
 Now as a simple assumption let us consider a particular distribution say a
delta function in the corresponding energy space:
\begin{equation}\label{ansatz}
f(p)d^3p=F(E)dE=F_0\delta(E-E_0)
\end{equation}
in which $E_0$ stands for the relevant energy state of the degenerate
particles. Using the above ansatz in (\ref{dens}) and (\ref{T11})
yields:
\begin{equation}
\rho=F_0E_0
\end{equation}
\begin{equation}
P={1\over 3}(F_0-{m^2\over E_0}F_0)
\end{equation}
employing the widely used  barotropic form of the equation of
state, i.e. $P=w\rho$, the above relations immediately lead to:
\begin{equation}
E_0={1 \over \sqrt{1-3w}}m,
\end{equation}
then one can proceed by requiring $w=-1$ as the condition for the
cosmic exponential expansion to obtain;
\begin{equation}
E_0= \frac{1}{2}m.
\end{equation}
Of course the reader notes that this exact result can be altered
by changing our previous assumption about the shape of the
distribution function, still it shows the capability of the
introduced conjecture to deal with the situation of the dark energy.

Introducing the ansatz (\ref{ansatz}) we are able to scrutinize
the forthcoming consequences. For the particle number density we
have:
\begin{equation}
n=\int{f(p)\mathrm{d^3}p}=\int{F(E)\mathrm{d}E}=\int{F_0\delta(E-{m\over
2})\mathrm{d}E}=F_0
\end{equation}
It is of course possible to evaluate the above integration by
putting $F_0=\frac{g}{h^3}m^3c^3$ based on dimensional reasons.
Then according to $n={\rho/m}$ we have;
\begin{equation}\label{result}
\rho = \int{EF_0\delta(E-m/2)\mathrm{d}E}=\frac{m}{2}F_0 =
\frac{gm^4c^3}{2h^3}.
\end{equation}
Then if our ansatz is to be responsible for dark energy, we can
equate the above result with the measured amount of the current
dark energy density to obtain an estimate for the mass of the
particles involved in producing the desired energy. To do this, we
set the density to (\ref{result}) and the dark matter density
$\rho_\Lambda\approx 6.91\times10^{-27} Kg/m^3$ \cite{0} which
leads to:
\begin{equation}
m\thicksim 0.001 eV
\end{equation}
The only known particles near to this mass scale are neutrinos \cite{5}, although we
do not insist on this coincidence at this stage.

 The second possibility $E<0$ leads to $\rho<0$, which may be
 employed as a source for the AdS spacetime. Though the negative energy
 sounds exotic but it appears again in Dirac's electron
 sea postulate.  Since here we are
 interested in dark energy with $\rho>0$, we do not discuss this
 case any further.

 As we mentioned earlier in this letter, our treatment requires imaginary momentum for the dark fluid particles.
 The most straightforward way to satisfy our requirement
(imaginary momentum) might be considering the square mass to be
negative or equivalently the mass itself to be pure imaginary.
Putting such an unusual mass term in Lagrangian results in an
unstable situation and the fields defined by such Lagrangian are
called tachyonic fields as first noted by Gerald Feinberg in 1967
\cite{2}. This paper was originally written to propose faster than
light particles in field theory, although afterward researches
declined this interpretation. The case $m^2<0$ reappeared as part
of the symmetry breaking Lagrangians and the Higgs field mechanism
\cite{8}. This scenario is the most explored scenario and it is
well known that an unstable potential with $m^2<0$ can bring about
the slow-roll mechanism and consequently the exponential expansion
of the universe during inflationary era.
 Tachyons, due to their formal definition, could not provide us the desired imaginary momentum.
although faster than light speed with respect to special
relativity seems the first candidate for our purposal due to
existing the $\gamma\equiv(1 - v^2/c^2 )^{-1/2}$ factor in
calculating the momentum, but the requirement of having a real
energy leads to an imaginary mass and the momentum again becomes
real \cite{2}. We therefore abandon the tachyonic possibility from
now on.

 The other possibility comes from non-relativistic
quantum mechanics, where particles demonstrate imaginary momentum
while tunnelling a barrier \cite{3}. We summarize that  having a
dominant field which is passing through a forbidden energy level (which is
less than the particle's rest energy)
via quantum tunnelling, can form a dark energy fluid and thus leading to
an accelerated expansion, a mechanism which
resembles the eternal inflation scenario.

\end{document}